# WISE/NEOWISE observations of Active Bodies in the Main Belt


James M. Bauer[1,2], A. K. Mainzer[1], Tommy Grav[3], Russell G. Walker[4], Joseph R. Masiero[1], Erin K. Blauvelt[1], Robert S. McMillan[5], Yan R. Fernández[6], Karen J. Meech[7,8], Carey M. Lisse[9] , Roc M. Cutri[2], John W. Dailey[2], David J. Tholen[7], Timm Riesen[7,8], Laurie Urban[7], Alain Khayat[7], George Pearman[2], James V. Scotti[5], Emily Kramer[6], De'Andre Cherry[1],Thomas Gautier[1], Stephanie Gomillion[1], Jessica Watkins[1], Edward L. Wright[10], and the WISE Team

[1] Jet Propulsion Laboratory, California Institute of Technology, 4800 Oak Grove Drive, MS 183-401, Pasadena, CA 91109 (email: bauer@scn.jpl.nasa.gov)

[2]Infrared Processing and Analysis Center, California Institute of Technology, Pasadena, CA 91125

[3]Planetary Science Institute, 1700 East Fort Lowell, Suite 106, Tucson, AZ 85719-2395

[4]Monterey Institute for Research in Astronomy, 200 Eighth Street, Marina, CA 93933

[5]Lunar and Planetary Laboratory, University of Arizona, 1629 East University Blvd., Kuiper Space Science Bldg. #92, Tucson, AZ 85721-0092

[6]Department of Physics, University of Central Florida, 4000 Central Florida Blvd., P.S. Building, Orlando, FL 32816-2385

[7]Institute for Astronomy, University of Hawaii, 2680 Woodlawn Dr., Manoa, HI 96822

[8]NASA Astrobiology Institute, University of Hawaii, Manoa, HI 96822

[9]Applied Physics Laboratory, Johns Hopkins University, 11100 Johns Hopkins Road Laurel, MD 20723-6099

[10]Department of Physics and Astronomy, University of California, PO Box 91547, Los Angeles, CA 90095-1547


*Short Title: WISE IR observations of* AMBO*s.*





## Abstract


We report results based on mid-infrared photometry of 5 active main belt objects (AMBOs) detected by the Wide-field Infrared Survey Explorer (WISE) spacecraft. Four of these bodies, P/2010 R2 (La Sagra), 133P/Elst-Pizarro, (596) Scheila, and 176P/LINEAR, showed no signs of activity at the time of the observations, allowing the WISE detections to place firm constraints on their diameters and albedos. Geometric albedos were in the range of a few percent, and on the order of other measured comet nuclei. P/2010 A2 was observed on April 2-3, 2010, three months after its peak activity. Photometry of the coma at 12 and 22 μm combined with ground-based visible-wavelength measurements provides constraints on the dust particle mass distribution (PMD), dlogn/dlogm, yielding power-law slope values of $\alpha = -0.5 +/- 0.1$. This PMD is considerably more shallow than that found for other comets, in particular inbound particle fluence during the Stardust encounter of comet 81P/Wild 2. It is similar to the PMD seen for 9P/Tempel 1 in the immediate aftermath of the Deep Impact experiment. Upper limits for $CO_2$ & CO production are also provided for each AMBO and compared with revised production numbers for WISE observations of 103P/Hartley 2.


## Introduction

Until recently no main belt asteroids were ever seen to exhibit dust ejection or cometary activity. However, in 1996 the discovery of activity from 133P/Elst-Pizarro (Elst *et al.* 1996), and later in multiple other bodies (e.g. Read *et al.* 2005, Hsieh & Jewitt 2006, Hsieh 2007), it became clear that a subset of the bodies within the Main Belt exhibited coma, some at regular intervals in their orbit, with the largest activity occurring near their





perihelion (cf. Hsieh et al 2010). As of late 2011, 7 bodies in the main belt have been identified as main belt comets (i.e. asteroids with associated dust tails, henceforth referred to as active main belt objects, or AMBOs). Their activity and optical qualities have been well-studied (cf. Hsieh *et al.* 2010, Hsieh *et al.* 2011a), but infrared measurements have been reported for only three bodies; Scheila (Tedesco et al. 2002 & 2004), and 176P and 133P (Hsieh et al. 2009). Here we present data taken in the thermal infrared by the Wide-field Infrared Survey Explorer mission (WISE; Wright *et al.* 2010) of 5 of these bodies, four of which appeared as point sources, and we provide measurements of their sizes and corresponding albedos. We also present our analysis of the dust surrounding the AMBO P/2010 A2 (LINEAR). Upper limits of $CO_2$ production are also presented and compared with recomputed values of $CO_2$ production for 103P/Hartley 2 that supersede previously published rates based on WISE fluxes.

The WISE mission surveyed the sky at four mid-IR wavelengths simultaneously, 3.4 μm (W1), 4.6 μm (W2), 12 μm (W3) and 22 μm (W4), with approximately one hundred times improved sensitivity over the Infrared Astronomical Satellite (IRAS) mission (Wright *et al.* 2010). The field of view for each exposure was 47x47 arcmin. Over 99% of the sky was covered with multiple exposures, averaging 10 per sky region, but varying in density as a function of ecliptic latitude. On the ecliptic, the minimum number of exposures per sky region was 8. As part of an enhancement to the WISE data processing system called "NEOWISE", the WISE Moving Object Processing Software (WMOPS) was developed to find solar system bodies in the WISE images (Dailey *et al.* 2010, Mainzer *et al.* 2011a). WMOPS successfully found a wide array of primitive bodies,





including Near-Earth Objects (NEOs), main belt asteroids, comets, Trojans, and Centaurs. By the end of the spacecraft mission, NEOWISE identified more than 157,000 small bodies, including 123 comets (Mainzer *et al.* 2011a). These infrared observations are useful for determining solid body size and albedo distributions, and thermo-physical properties such as thermal inertia, the magnitude of non-gravitational forces, and surface roughness (Mainzer *et al.* 2011b, 2011c). The subset of these bodies exhibiting cometary activity require special treatment in the interpretation of such observations, owing to the material surrounding the solid nucleus, i.e. the contribution to the IR flux from the gas and dust, and the variable nature of the observed brightness of the object attributable to outbursts. These IR imaging data provide unique opportunities to characterize four main components of comets: the dust and gas comae, the nucleus, and the extended dust trail. Here we apply these techniques (Bauer *et al.* 2011) to the 5 AMBOs detected by WISE: (596) Scheila (hereafter Scheila) , 133P/Elst-Pizarro (133P), 176P/ LINEAR (176P) , P/2010 R2 (La Sagra) (P/2010 R2), and P/2010 A2 (LINEAR) (P/2010 A2). The remaining two known AMBOs, 238P/Read and P/2008 R1 (Garradd) were not detected in any band.

## Observations & Analysis

The WISE spacecraft surveyed the sky as its terminator-following geocentric polar orbit progressed about 1 degree of ecliptic longitude per day. Regular survey operations commenced on 2010 Jan. 14 (MJD 55210), imaging the sky simultaneously in all four bands until the cryogen was depleted in the secondary tank on 2010 Aug 5 (MJD 55413). The survey then entered a three-band (W1-W3) phase that lasted through 2010 Sep 30





(MJD 55469). The final phase, the post-cryogenic mission with W1 and W2, lasted from 2010 Oct 1 through 2011 Jan 31, (MJD 55592; cf. Cutri *et al.* 2011). All photometric data of detected objects presented here were obtained during the first phase, except for some last-phase observations of Scheila. Some of the imaging data presented here includes observations of objects in the post-cryogenic mission phase.

During the fully cryogenic portion of the mission, simultaneous exposures in the four WISE bands were taken once every 11 sec, with exposure durations of 8.8 sec in W3 and W4, and 7.7 sec in W1 and W2 (Wright *et al.* 2010). The number of exposures acquired on moving objects varied depending on the location of the object on the sky (especially its ecliptic latitude), the toggle pattern of the survey employed to avoid imaging the Moon, and the relative motion of the object with respect to the progression of the survey (Mainzer *et al.* 2011a, Cutri *et al.* 2011). Note that WISE may have observed an object while it was in different patches of the sky, i.e. when several weeks or months had passed since the previous exposure; henceforth, we refer to the series of exposures containing the AMBO in the same region of sky as a "visit".

Table 1 lists the exposures for each of the 5 detected AMBOs, as well as their mean viewing phase angles, and heliocentric and observer distances during each visit. With two separate visits, (596) Scheila had the greatest coverage depth in W1 and W2, while P/2010 R2 had the greatest depth of W3 and W4 coverage. The last 4 entries of Table 1 also show the frames that covered the predicted locations of 238P and P/2008 R1 (Garradd), which were not detected in either the individual or stacked frames, but from which we derive upper limits to their fluxes (see Table 2).





**Table 1: Mid-IR Observations of Known Active Main Belt Asteroids**

| Object[a] | N[a] | R [a] (AU) | Δ [a] (AU) | α [a] (°) | Coma ? [a] | Stacked ? [a] | Comments[a] |
|---|---|---|---|---|---|---|---|
| (596) Scheila | 10 | 3.93 | 3.14 | 16.8 | No | No | -- |
| MJD[a] Start Times: 55242.35929406, 55242.49159826, 55242.62390243, 55242.69011816, 55242.82242237, 55242.95472657, 55242.95485389, 55243.08715810, 55243.21946226, 55243.35176641 |||||||| 
| 133P | 13 | 3.67 | 3.42 | 15.6 | No | Yes | -- |
| MJD Start Times: 55271.99996748, 55272.13239883, 55272.26470277, 55272.39700681, 55272.46309514, 55272.46322245, 55272.52931076, 55272.59552641, 55272.66161475, 55272.72783039, 55272.79391878, 55272.86013443, 55272.99243847, 55273.12474251 |||||||| 
| P/2010 A2 | 16 | 2.07 | 1.73 | 28.8 | Yes | Yes | -- |
| MJD Start Times: 55288.27578843, 55288.40809252, 55288.54039657, 55288.67270062, 55288.80500469, 55288.87122037, 55288.93730874, 55289.00352447, 55289.06961284, 55289.13582852, 55289.20191689, 55289.26813252, 55289.40043657, 55289.53274063, 55289.66504468, 55289.79734873 |||||||| 
| 176P | 16 | 3.15 | 2.99 | 18.6 | No | Yes | -- |
| MJD Start Times: 55309.28058017, 55309.41288442, 55309.54518866, 55309.67749290, 55309.67762031, 55309.80979715, 55309.80992451, 55309.87601297, 55309.94210145, 55309.94222876, 55310.00831722, 55310.07453300, 55310.14062147, 55310.27292571, 55310.40523000, 55310.53753425 |||||||| 
| P/2010 R2 | 19 | 2.62 | 2.40 | 22.8 | Yes | Yes | -- |
| MJD Start Times: 55356.64184157, 55356.77414537, 55356.90644912, 55357.03875296, 55357.17092939, 55357.17105676, 55357.23714494, 55357.30323309, 55357.36944865, 55357.43553689, 55357.50175322, 55357.56784223, 55357.63405781, 55357.76636159, 55357.89853814, 55357.89866545, 55358.03084197, 55358.03096933, 55358.16314579 |||||||| 
| (596) Scheila | 14 | 3.15 | 2.98 | 18.3 | No | No | -- |
| MJD Start Times: 55510.88545743, 55511.01763413, 55511.14993825, 55511.34833072, 55511.41441912, 55511.41454644, 55511.48063478, 55511.54672802, 55511.61281640, 55511.61294377, 55511.67903215, 55511.87742472, 55512.00972890, 55512.14190566 |||||||| 
| 238P | 32 | 2.86 | 2.69 | 20.7 | -- | -- | *No Detection* |
| MJD Start Times: 55320.52708031, 55320.65938453, 55320.79168874, 55320.92399296, 55321.05629718, 55321.12251292, 55321.18860130, 55321.25481710, 55321.32090552, 55321.38712131, 55321.45320973, 55321.51942547, 55321.65172964, 55321.78403387, 55323.57007663, 55323.70238080, 55323.70250817, 55323.83468502, 55323.83481234, 55323.96711712, 55324.09942130, 55324.16550962, 55324.23172537, 55324.29781380, 55324.36402954, 55324.43011796, 55324.49633370, 55324.56242208, 55324.69472630, 55324.82703043, 55324.95933464 |||||||| 
| P/2008 R1 | 13 | 3.46 | 3.27 | 16.6 | -- | -- | *No Detection* |
| MJD Start Times: 55263.26917722, 55263.40148125, 55263.53378520, 55263.66608919, 55263.73230484, 55263.79839317, 55263.79852053, 55263.86460888, 55263.93082453, 55263.99691291, 55264.12921690, 55264.26164819, 55264.39395223 |||||||| 
| 238P | 15 | 2.46 | 2.17 | 23.6 | -- | -- | *No Detection* |
| MJD Start Times: 55502.32299741, 55502.45530140, 55502.58747807, 55502.71978206, 55502.78587042, 55502.85208605, 55502.91817441, 55502.98426272, 55502.98439003, 55503.05047840, 55503.11656671, 55503.18278234, 55503.31495902, 55503.44726300, 55503.57956699 |||||||| 
| P/2008 R1 | 14 | 3.66 | 3.53 | 15.7 | -- | -- | *No Detection* |
| MJD Start Times: 55518.82066059, 55518.95296481, 55519.08514168, 55519.08526905, 55519.21744591, 55519.28353439, 55519.28366170, 55519.34975019, 55519.41583862, 55519.48205441, 55519.54814285, 55519.68044707, 55519.81262398, 55519.94492822 |||||||| 





[a] Each object is listed per visit (see text). N refers to the number of exposures, R is the heliocentric distance of the AMBO in AU, $\Delta$ is the observer distance in AU, $\alpha$ is the phase angle in degrees. The "Coma?" column refers to whether there is apparent coma in the images. "Stacked?" indicates whether the analysis was performed on a stacked (co-added) image of the N exposures, to increase the signal from the AMBO, or whether each exposure was individually analyzed; *No Detection* indicates there was no detection in the stacked or individual images. MJD, i.e. the Modified Julian Date, is defined as JD-2400000.5. The range of dates for each listed visit is as follows (all dates are 2010): (596) Scheila (1st visit) – Feb 15, 05:26:52 to Feb 16, 08:26:33; 133P – Mar 16, 23:59:57 to Mar 18, 02:59:38; P/2010 A2 – Apr 02, 06:37:08 to Apr 03, 19:08:11; 176P – Apr 23, 06:44:02 to Apr 24, 12:54:03; P/2010 R2 – Jun 09, 15:24:15 to Jun 11, 03:54:56; (596) Scheila (2nd visit) – Nov 10, 21:15:04 to Nov 12, 03:24:21; 238P (1st visit) – May 04, 12:39:00 to May 08, 23:01:27; P/2008 R1 (1st visit) – Mar 08, 06:27:37 to Mar 09, 09:27:17; 238P (2nd visit) – Nov 02, 07:45:07 to Nov 03, 13:54:35; P/2008 R1 (2nd visit) – Nov 18, 19:41:45 to Nov 19, 22:40:42.

The WISE image data were processed using the scan/frame pipeline that applied instrumental, photometric, and astrometric calibrations (Cutri *et al.* 2011). Each of our five objects was observed more than the average 10 times owing to the object's (prograde) motion being in the direction of the scan progression of the spacecraft. WISE covered all ecliptic latitudes each day in two narrow longitude bands at 95 +/- 2 degrees ahead of the Sun and 90 +/- 2 degrees behind the Sun, and used the spacecraft orbital motion around the Sun to scan this band across all ecliptic longitudes over 6 months. P/2010 A2 had the maximum sky-plane motion of the 5 AMBOs with a rate of 42 to 61 arcsec/hr ("moving" mostly via the reflex motion due to parallax and the spacecraft's orbital motion). At most, the apparent motion created a blurring of ~0.15 arcsec, an insignificant factor in the imaging, as the blur was far smaller than the pixel scale in the shortest wavelengths (2.75 arcsec/pixel in W1, W2, and W3; 5.5 arcsec/pixel in 2×2-





binned W4 images; Wright *et al.* 2010), and the PSF FWHM (6.5 arcsec in the three shortest wavelength bands). Scheila and P/2010 A2 were detected in individual exposures, with sufficient signal to be detected with NEOWISE moving object pipeline software (WMOPS; Mainzer *et al.* 2011a). The images of P/2010 A2 were stacked in order to increase the signal to noise of the dust tail; the images for P/2010 R2, 176P, and 133P were stacked to increase any signal from these bodies that were present, and the detections of the objects were identified only in the stacked images. The images for the AMBOs were shifted to match the sky-motion rates of each object as predicted by JPL's Horizon's ephemeris service (http://ssd.jpl.nasa.gov), and co-added using the "A WISE Astronomical Image Co-adder" (AWAIC) algorithm as described in Masci & Fowler (2009). All images were stacked in this manner for each corresponding visit to an object to conduct the photometric and morphological analyses.

For Scheila, 176P, 133P, and P/2010 R2, the image profiles including those of the stacked images were consistent with point-spread functions (PSFs), while the profile of P/2010 A2 was not consistent with a PSF, owing to the presence of coma. Special consideration was given to P/2010 R2 and to the December visit of Scheila (detected in the W1 and W2 bands only), owing to the fact that the WISE observations were relatively close in time to reported activity. No coma signal was apparent in the images (see Figure 1), nor was any significant coma signal found for P/2010 R2 or Scheila when surface brightness profile analysis techniques were applied (cf. Bauer *et al.* 2008). The surface brightness profiles (SBPs) shown in Figure 2 were generated from the W3 stacked image for P/2010 R2, and from the W2 stacked image for Scheila during the WISE spacecraft's second visit to the AMBO, i.e. in the post-cryo mission data. The PSFs and AMBO SBPs





were sampled out to 30 arcseconds, well into the region where each object's SBP wings reach into the local background.  Analysis also showed that fluxes for Scheila were consistent with the flux values derived from the February 2010 visit data when re-scaled for distance and phase angle (IAU phase parameter G=0.08; Bowell *et al.* 1989), suggesting that there was no surrounding dust or out-gassing.

**Table 2: Total Fluxes in mJy**

| Object | W1 (3.5 $\mu$m) | W2 (4.6 $\mu$m) | W3 (12 $\mu$m) | W4 (22 $\mu$m) | Log($Q_{CO2}/Q_{co}$) |
|---|---|---|---|---|---|
| Scheila *cryo* | 8.7 +/- 0.5 | 17 +/- 1 | 2700 +/- 300 | 6800 +/- 800 | -- |
| Scheila *post-cryo* | 11.0 +/- 0.6 | 30 +/- 2 | -- | -- | -- |
| 133P | < 0.03 | < 0.05 | 1.3 / - 0.2 | 4.0 +/- 0.6 | < 25.6, < 26.6 |
| 176P | < 0.02 | < 0.06 | 3.2 +/- 0.2 | 8.0 +/- 0.8 | < 25.5, < 26.5 |
| P/2010 R2 | < 0.01 | < 0.03 | 1.33 +/- 0.07 | 5.2 +/- 0.5 | < 25.0, < 26.0 |
| P/2010 A2 | < 0.09 | < 0.11 | 57 +/- 9 | 124 +/- 21 | < 25.3, < 26.3 |
| 238P | < 0.02 | < 0.06 | < 0.2 | < 0.5 | < 25.4, < 26.4 |
| P/2008 R1 | < 0.04 | < 0.06 | < 0.2 | < 0.6 | < 25.6, < 26.6 |

Aperture photometry was performed on the stacked images of 176P, 133P, & P/2010 R2 for aperture radius values of 11 and 22 arcsec, the aperture sizes necessary to obtain the full signal from W3 and W4, the poorest resolution WISE bands, while pipeline-extracted magnitudes were used for the thermal fits of Scheila. The counts were converted to fluxes using the band-appropriate magnitude zero-points and 0[th] magnitude flux values provided in Wright *et al.* (2010), and an iterative fitting to a black-body curve was conducted on the two long-wavelength bands to determine the appropriate color correction as listed in the same. The extracted magnitudes were then converted to fluxes (Wright *et al.* 2010, and Mainzer *et al.* 2011b) and are listed in Table 2. Proper aperture corrections are required for accurate photometry (Cutri *et al.* 2011), in addition to the color corrections mentioned above. With these corrections, the derived magnitudes are equivalent to the





profile-derived magnitudes providing there are no artifacts, saturation, or confusion with other sources in the images of the objects. Note that the profile magnitudes for Scheila allowed for a more accurate photometric magnitude in W3, since the core of the image was saturated for this object (cf. Mainzer *et al.* 2011b). Table 2 also lists the 1-σ upper limits of fluxes in W1 and W2 for the remaining AMBOs with PSF-like profiles that lack detections in these bands. Corresponding $CO_2$ and CO upper limits are also provided in units of log mol sec$^{-1}$, based on analysis outlined in Pittichova *et al.* (2008) and Bauer *et al.* (2011). The values listed assume a single source species for the observed upper limit in W2. Note that in the course of the analysis, the $CO_2$ and CO values were re-computed for 103P/Hartley 2, and were found to be off by a factor of 17 as reported in Bauer *et al.* (2011) when using a higher precision code. The corrected column densities for $CO_2$ and CO are 3.0 ($\pm2.1$) $\times$ 10$^{11}$ and 3($\pm2$) $\times$ 10$^{10}$ m$^{-2}$ respectively. Production values therein should also be corrected as 6.0 ($\pm2.0$) $\times$ 10$^{25}$ and 6.5 ($\pm$ 2.2) $\times$ 10$^{24}$ mol sec$^{-1}$ for $CO_2$ and CO, respectively. The relative production rates for $CO_2$ as compared to the predicted level of water production, then, are on the order of 20%, rather than the few percent stated in Bauer *et al.* (2011).  The AMBO production upper limits compare with the103P $CO_2$ production rate of 25.78 in log units. These AMBO production upper limits are weak constraints, owing to the greater distance of the AMBOs from WISE,  relative to 103P, at each object's times of observation.  The AMBO flux upper limits are on the order of the expected confusion limits for each band (Cutri *et al.* 2011), as are the upper limits derived from the stacked images of 238P and P/2008 R1. Using the thermal fit parameters for 176P, we find 3-σ upper limits of 2.0 km and 1.8 km (or 1-σ limits of 1.2 and 1.1 km) for 238P and P/2008 R1, respectively. We list the 1-σ upper limits in Table





3, and note that the limit is consistent with the 238P diameter estimate of 0.8 km by Hsieh et al. 2011b and that the region of sky containing the AMBO was observed by WISE during the reported time of inactivity.

To constrain the albedo, $H_V$ values were obtained from the literature for 176P (Hsieh *et al.* 2011a), 133P (Hsieh *et al.* 2010), and Scheila (cf. Tedesco *et al.* 2004). 176P and 133P have well-characterized rotation periods of 22.2 and 3.47 hours (Hsieh et al. 2009, Hseih et al. 2010), respectively. The WISE images sampled ~1.3 rotations for 176P, and ~7.7 rotations for 133P. The closest match to the WISE data's sampling cadence and the rotation period is for 133P. WISE orbits once every ~95 minutes (or two orbits every 3.1753 +/- 0.0003 hours as indicated by the times in Table 1), while the reported rotation period for 133P is 3.47(1) hours. Each point is sampled ~8% of the period off its previous nearest sample point in rotational phase. For 13 visits (not counting the double-sample where 133P's image fell in the frame overlap region near time MJD 55272.463), this covers 96% of the range of phase space at roughly 8% intervals. Assuming there is a small bias in the selection of data points, the reported amplitude is ~0.1 magnitudes, which is not far off from the fit uncertainty reported in Table 3. Note that even with the < 1% rotation period uncertainty reported in Hsieh et al. 2010, it is not possible to extrapolate the rotational phase from the reported observations of 2007. For each visit of Scheila, the span of observations, ~24 and 30 hours, was well over a full rotation period of 15 +/- 5 hours (cf. Warner *et al.* 2009). For matching visual-band observations of P/2010 A2, we reduced Spacewatch data from the 0.9m Steward Observatory telescope (cf. Larsen *et al.* 2007) for the night of 2010 Mar 15, preceding the WISE observations by about 20 days. Using on-frame sources listed in the USNO-1B catalog (Monet *et al.*





2003) to conduct relative photometry we found a V magnitude of 18.3 +/- 0.4. We compared the derived values with those from Jewitt *et al.* (2011a) and found that the magnitudes matched those taken within a day of our 2010 Mar 15 observations. Jewitt *et al.* (2011a) also listed reported magnitudes within hours of the WISE April observations, with total magnitude $H_v = 14.4$, and we used this value to constrain our visual-band brightness for P/2010 A2. The only publically available values for P/2010 R2 were from the Minor Planet Cener (http://www.minorplanetcenter.net), which listed the $H_v$ value as 15.1. However, project NEAT observe the predicted position of the AMBO on February 9, 2002 from the Palomar 48-inch, when the predicted R-band brightness was 20.1. The three images were provided by the NEAT archive project (cf. Lawrence *et al.* 2009), and were stacked matching the predicted sky motion of the AMBO. A signal-to-noise ratio of 4 was achieved down to $m_R = 20.5$, and no corresponding source was found near the predicted location. Therefore we instead used an $H_V$ for P/2010 R2 derived from observations taken at the University of Hawaii 2.2 meter telescope on Mauna Kea, HI on August 1 and 9, 2011 (UT) in place of the MPC's listed value. These observations yielded an estimated magnitude of 23.8 +/- 0.1 in R-band (R=3.0 AU, $\Delta$=3.5 AU, $\alpha$=16°), corresponding to the $H_v \sim 18.0$, assuming an IAU phase slope parameter of G=0, and near-solar colors. This was the best available estimate; those at the MPC were based on observations obtained close to the time of apparent onset of activity. The measured colors of 133P (Hsieh et al. 2010), 176P (Hsieh et al. 2009), and (596) Scheila (pre-outburst; Tedesco 1995 and Yang & Hsieh 2011) are near-solar, within 0.06 magnitudes, and G-parameters are in the range of -0.1 to 0.1. Assuming an offset in both values (G=0.1 and V-R=0.42), the magnitude offset would be ~0.12, or approximately 12% in brightness, in





both cases, considerably less than the uncertainty in albedo listed in Table 3, and on the order of the photometric uncertainty.

## Discussion

Using a NEATM model (Harris *et al.* 1998, Mainzer *et al.* 2011b) thermal fits were conducted on the photometric fluxes of the AMBOs without apparent coma (Figure 3); the fits are summarized in Table 3. We present here the fit results and uncertainties, while it should be noted that, owing to the uncertainty in the absolute calibration of the WISE thermal bands, there is an additional ~10% uncertainty in the derived diameter values (Mainzer *et al.* 2011b & c). Hsieh et al. 2009 report sizes for 133P and 176P based on the Spitzer Space Telescope MIPS 24 μm signal alone. The sampling cadence of the MIPS data consisted of three exposures taken over an 8 minute interval for 133P and two exposures spaced 5 hours apart for 176P. As discussed previously, the WISE observations consisted of 13 and 16 visits for 133P and 176P respectively, spaced at 1.59 hour intervals and with a more complete sampling of each body's rotational phase. WISE observed both bodies far from the heliocentric distances about their perihelion where their activity was previously reported (Hsieh et al. 2010 & Hsieh et al. 2011a), and comparable to the distances of the Spitzer Space Telescope observations reported in Hsieh et al. 2009. The WISE data had two thermal channels at 12 and 22 μm, which allowed for a fit with η as a free parameter, and the fits converged to solutions of η near 0.8. Considering these factors, the WISE results of size and albedo compare well with the Hsieh et al. 2009 results for the fixed η ~ 0.8. Converting the Hsieh et al. 2009 values of $p_R$ to $p_V$, we derive for 133P an albedo of $p_V = 0.04$ +/- 0.01, and for 176P, $p_V = 0.05$ +/- 0.01, which overlap with our values in Table 3. Hsieh et al. (2009) report sizes ranging





from 3.34 - 3.56 km for 133P and 3.44 - 4.08 km for 176P using $\eta = 0.8$ calculations, which overlap our derived values and uncertainties, although our sample likely falls closer to the mean, based on the WISE imaging cadence for each object.

We used a default value of 2 for the ratio of $p_{IR}/p_V$ in the thermal fit results. Note that the fits to this ratio are only loosely constrained by the W3 and W4 signal for 133P, 176P, and P/2010 R2, and so the value of $p_{IR}/p_V$ is close to the default value used. For Scheila, however, $p_{IR}/p_V$ was strongly constrained by the additional W1 and W2 signal, so that the ratio of 2 is a firmly fitted result, but not unlike what has been found for the WISE data for other redder (V-R > 0.36) main belt objects (cf. Mainzer *et al.* 2011b), as in the case of those in the Scheila D-type spectral class (Yang & Hsieh 2011).

**Table 3:  Object Nucleus Thermal Fits**

| Object | Diameter (km) | $p_v$ | $p_{ir}$ | $\eta$ |
|--------|---------------|-------|----------|--------|
| Scheila | 118 +/- 6 | 0.04 +/- 0.004 | 0.08 +/- 0.03 | 0.83 +/- 0.03 |
| 133P | 3.2 +/- 0.2 | 0.06 +/- 0.02 | 0.12 +/- 0.03 | 0.8 +/- 0.1 |
| 176P | 3.5 +/- 0.1 | 0.07 +/- 0.03 | 0.15 +/- 0.05 | 0.8 +/- 0.1 |
| P/2010 R2 | 2.8 +/- 0.3 | 0.01 +/- 0.01 | 0.02 +/- 0.02 | 1.9 +/- 0.3 |
| P/2010 R2 | 1.6 +/- 0.3 | 0.03 +/- 0.02 | 0.05 +/- 0.03 | *Fixed at* 0.8 |
| 238P | $\leq$1.2 | $\geq$ 0.03 | -- | *Fixed at* 0.8 |
| P/2008 R1 | $\leq$1.1 | $\geq$ 0.04 | -- | *Fixed at* 0.8 |

Physical parameters derived from the WISE data differ between the dust of active coma and solid nucleus surfaces. The general properties of the WISE data have been discussed in detail by Cutri *et al.* (2011), and the performance of thermal models applied to WISE observations of solid bodies is described in Mainzer *et al.* (2011b,c).  Methods used in the analysis of the coma dust particles of P/2010 A2 were similar to these introduced in Bauer *et al.* (2007, 2008, & 2011). Analysis of the flux of coma constrains the dust





particle size distribution and the quantity of CO and $CO_2$ emitted by the comet. The IR fluxes for solid nuclei provide constraints on the size of the comet and, when accompanied by shorter (non-thermal) wavelength data, constrain surface albedo values as well.

P/2010 A2 was the only AMBO in our sample to exhibit an apparent dust tail while WISE observed the body. As the coma dominated the signal, no special extraction of the nucleus signal was possible. Thermal fits to the coma of P/2010 A2 were conducted using a Planck function (Figure 4), similar to the analysis conducted on 103P/Hartley 2 (Bauer *et al.* 2011). The contribution of the nucleus was not removed in this case, as the signal from the nucleus, predicted to be ~120 meters across (Jewitt *et al.* 2010, Snodgrass *et al.* 2010), was less than 2% of the total signal, and WISE was unable to resolve it separately from the dust signal. The best thermal fit to a Planck function is shown in Figure 4, and yields a higher temperature (238K) than the expected black-body temperature (200K) for P/2010 A2's solar distance. One possible explanation for this is that the coma is dominated by very small (a few μm) grains that are highly absorbing in the optical but poorly emitting in the far-IR. Small grains are also likely to evince silicate emission bands (cf. Kolokolova *et al.* 2004), in which case the W3 excess is ~30% of the signal, provided the grains are nearly the same temperature as a black body. Another more likely possibility, that the grains are not isothermal emitters, which may be the case for large (> 1cm) grain sizes, is supported on dynamical grounds from the imagery of Jewitt *et al.* 2011a, Snodgrass *et al.* 2010, and Moreno *et al.* 2010. Since, as in Figure 1, the P/2010 A2 dust follows the P/2010 A2 orbit closely, larger (> 100 μm) grains are likely dominant. While very near its perihelion (perihelion was on 2009 Dec 4 at a heliocentric





distance of 2.0055 AU, outburst discovery heliocentric distance on 2009 Dec. 15 was 2.0061 AU), the activity in this AMBO is believed to have been initiated by an impact event (Jewitt *et al.* 2010), though some uncertainty may remain (Moreno *et al.* 2010). As with all the AMBOs except for Scheila, no W1 or W2 signal was observed in either the individual or stacked images of P/2010 A2, therefore no estimates of CO or $CO_2$ production could be derived, but only upper limits. The near-simultaneous Hubble Space Telescope (HST) photometry (within hours of the WISE observations; Jewitt *et al.* 2011a) provides constraints on the reflected-light dust signal. Using a method identical to that described in Bauer *et al.* (2011), scaling by the effective projected area of the dust in the coma, we were able to find the number of dust particles ($n_{dust}$) contributing to the signal in the photometry aperture for each wavelength interval (Table 4). Assuming a new particle size in each band similar to the wavelength scale, and subtracting the contribution from each of the preceding band-centered wavelengths, starting with the longest (W4), we derived a particle mass distribution (PMD; d*log*n/d*log*m, with m as the particle mass with constant density $\rho$=1 g cm$^{-3}$) as shown in Figure 5. A log-log slope was fit to the result, for a comparison with other comets, and found to be $\alpha$ = -0.5 +/- 0.1 (note that the aforementioned W3 excess does not significantly affect this fit), considerably more shallow than most active comets, which more commonly fall within the $\alpha$ = -0.8 to -1.2 range (Lisse et al. 1998, 2004, 2007; Fulle *et al.* 2004; for comparison, divide the values therein by 3. The particle size distribution slope parameter values computed therein are scaled with respect to size, rather than mass, resulting in a factor of 3 steeper than ours). Our slope value is instead closer to the slope seen in the immediate post-impact of the Deep Impact experiment on 9P/Tempel 1 (cf. Bauer *et al.*





2007), consistent with a PMD caused by an impact-driven event. However, an alternative explanation is more likely; a large number of the finer particles may have been driven away from the coma by the solar radiation pressure over the course of the 15 weeks since the first observed activity, or 6 - 11 months since the projected onset of activity (Jewitt *et al.* 2010, Snodgrass et al. 2010, Moreno et al. 2010). In that case, no new small particles would have been produced to replace those swept away, implying again a sudden outburst limited in duration, and not extending into the epoch of the WISE observations.

**Table 4:  Particle Mass Distribution**

| Quantity | R-band | 12μm | 22μm |
|---|---|---|---|
| $m_g$[kg] | $1.8 \times 10^{-16}$ | $8.1 \times 10^{-13}$ | $5.7 \times 10^{-12}$ |
| $n_g$ | $1.3 \times 10^{28}$ | $5.0 \times 10^{15}$ | $1.4 \times 10^{16}$ |

Objects 133P, 176P, P/2010 R2, and Scheila all demonstrate image morphologies that matched the observational stellar PSFs. 176P and 133P were not close to their perihelia, i.e. when the bodies were observed to have been the most active in the past (cf. Hsieh et al 2010 & 2011). However, P/2010 R2 was approaching its perihelion distance of 2.62 AU and was observed to be active 65 days after WISE observed the object. It is still possible that P/2010 R2 was undergoing low-level activity at the time of the WISE observations in late June. The fact that the best fit beaming parameter for P/2010 R2 is 1.9 is suggestive, though other fits with lower η values are still feasible (See Figure 3D). Fits with η in the range of  0.8-1.2, although poorer,  fall within the 95%  confidence level and produce geometric albedo values in the range of 2-3.4%, closer to the 4-6% values of other AMBOs, rather than the lower 1% value for the best fit. We listed the P/2010 R2 fit for the fixed-η value of 0.8 in Table 3 in addition to the free-η fit for





comparison with the other three AMBO fits, all of which yield η values near 0.8. For

comparison, Fernandez et al. (2011) find the mean η value in their Jupiter family comet

sample to be near 1.0, with a standard deviation of 0.1, significantly less than the high-η

best fit for P/2010 R2. The best-fit high η value implies the temperature may be cooler

than expected, and it may be cool enough to be explained by the presence of isothermal

dust grains. Alternatively, the fit to a Planck function shown in Figure 3E is elevated by

27K from the black-body temperature, which may be caused by the same phenomena

(abundant small or large grains or a pronounced silicate emission feature) discussed for

P/2010 A2, or alternatively caused by a signal dominated more by the nucleus. If so, the

nucleus size derived from the thermal fits would serve more as an upper limit, since

activity would likely enhance the IR flux. The fact that the object shows no sign of

activity in WISE data from surface brightness profile analysis could be at least partially

due to the large pixel scales of the WISE data. The use here of an optical magnitude

inferred from actual inactive data could have resulted in an underestimate the true optical

magnitude at the time of the WISE observations, leading to an underestimated albedo.

The limits of activity for Scheila are more firmly constrained in that the object size

derived from the thermal flux, and the corresponding albedo, match those found in the

literature (cf. Tedesco *et al.* 2004). Larson (2010) mentions a possibly star-like

appearance as late as late as Nov. 11, we provide true photometric and surface brightness

constraints on the activity beyond this date. Furthermore, the W1 data are consistent with

a lack of activity when they are corrected for distance and phase angle. For each visit in

February and November, the W1 signal is dominated by reflected light for distances





~3AU, and are equal with each other to within < 1% when corrected for heliocentric and observer distances listed in Table 1, and the observational phase angle change from 18.3 to 16.8 degrees, using an IAU slope parameter of G=0.08. Hence a clear and closer constraint is placed on the time of Scheila's outburst, within 21 days of the earliest reported activity when the AMBO was ~1.3 magnitudes in excess brightness on 2010 Dec 3 (Jewitt *et al.* 2011b, Bodewits *et al.* 2011).

The average AMBO albedo derived from WISE observations taken from these data, is 0.06, and the standard deviation in the sample is 0.02. This mean is consistent with other measurements of AMBO albedos (Hsieh *et al.* 2009) and is more or less consistent with measured comet reflectances which are ~0.04 (cf. Lamy *et al.* 2004). Objects 133P and 176P have been noted as having dynamical and physical similarities with the Themis family asteroids (cf. Hsieh et al. 2009), including similar albedo values. Our own albedo values for 133P (0.06 +/- 0.02) and 176P (0.07 +/- 0.03) affirm this comparison, and match the values WISE has measured for the subset of Themis members in Masiero et al. (2011), which has a mean of 0.07 +/- 0.02.

## Conclusions

WISE has managed to sample the majority of the known AMBOs in the thermal and mid-IR. One AMBO, P/2010 A2, was dominated by its dust-coma signal, while the others were not likely active at the time of their observations. From the observed fluxes we conclude the following:

- The thermal fits for P/2010 A2 yield higher temperature Planck functions than the black body temperature at the observed solar distance by 20% (38K), which can most readily be explained by large, non-isothermal grain dust. The slope of the





PMD, in units of d*log*n/d*log*m is -0.5 ± 0.1. The PMD of P/2010 A2, when fit

with a power law, is similar to that seen in an impulsive outburst, but most likely

indicates that the activity was over a finite window of time several months in the

past.

- The onset time of activity for Scheila is further constrained by our data to be

  within 21 days of the first observation of activity. The derived surface reflectance

  and diameter are consistent with literature values ($p_V$=0.04+/-0.008, D=115+/-6,

  cf. Tedesco *et al.* 2004). While the albedo and thermal inertia are entirely in line

  with canonical cometary values, the derived diameter is one of the largest values

  measured for a comet. Scheila is apparently larger than many of the outer planet's

  smaller moons and the median diameter of main belt asteroids (cf. Bottke et al.

  2005).

- AMBO nuclei albedos are consistent with measured comet albedos, i.e. are on the

  order of a few percent.

## Acknowledgements

This publication makes use of data products from the Wide-field Infrared Survey

Explore, which is a joint project of the University of California, Los Angeles, and the Jet

Propulsion Laboratory/California Institute of Technology, funded by the National

Aeronautics and Space Administration. This publication also makes use of data products

from NEOWISE, which is a project of the Jet Propulsion Laboratory/California Institute

of Technology, funded by the Planetary Science Division of the National Aeronautics and

Space Administration.  NEAT archive data was provided through NASA's Planetary





Mission Data Analysis Program. Observing time was allocated at Steward Observatory's 0.9m (Spacewatch) telescope on Kitt Peak. J. Bauer would also like to thank Drs. Hsieh and Jewitt for their valuable discussions regarding AMBOs. This material is based in part upon work supported by the NASA through the NASA Astrobiology Institute under Cooperative Agreement No. NNA09DA77A issued through the Office of Space Science.

**Figure Captions**

**Figure 1** A three-color composite image of AMBOs from the WISE data. The stacked 4.6, 11.6, and 22.1 μm images were mapped to blue, red, and green channels The AMBOs from left to right are (top row) P/2010 A2 and P/2010 R2, and (bottom row) Scheila, 176P, and 133P. The P/2010 A2 image is 9 arcmin across its bottom edge, while the others are 4.5 arcmins across. The P/2010 A2 panel shows the sky-projected anti-solar vector as indicated by the green dashed arrow, and the projected anti-velocity vector by the white dotted arrow. Note that while Scheila is saturated in W3, the profile photometry used in our analysis is still viable (Cutri *et al.* 2011).

**Figure 2** Surface brightness profiles for Scheila in W2 (top panel) and P/2010 R2 in W3 (bottom panel) sampled out to 24 arcseconds. The PSFs were constructed from a nearby bright star in the images for P/2010 R2. For Scheila, which had no nearby stellar counterpart of similar brightness, the comparison PSF was constructed from the array of synthetic PSFs available from the WSDS (Cutri *et al.* 2011), which oversample the PSF variability across the chip. The PSFs that were coadded to form the comparison PSF were appropriately selected based on





the pixel location of Scheila on each image. Error bars shown include the photometric uncertainties of the objects as well as the uncertainty in the background level. Note that owing to the considerably lower signal-to-noise-ratio for P/2010 R2, the SPB is more coarsely sampled, i.e. binned over twice the interval than that used for Scheila. Magnitude values are instrumental, based on image counts and default zero points (cf. Wright *et al.* 2010), uncorrected for color.

**Figure 3** The thermal fits for the AMBOs exhibiting PSF-shaped profiles: 133P (panel A), 176P (panel B), Scheila (Panel C) and P/2010 R2 (Panels D & E). Panels A-D show fits using the NEATM model appropriate for signal dominated by solid nuclei (Harris *et al.* 1998, Mainzer *et al.* 2011b), with beaming parameter values of $\eta=0.8$ (dotted lines), $\eta=1.0$ (dashed lines), and $\eta=1.2$ (dot-dahsed lines). Note that in Panel C, the fit to 596 converged freely to $\eta=0.8$ (see Tabel 3 for the best-fit parameters), and that for panel D we include the best-fit model of $\eta=1.9$ (dot-long-danshed line); the $\eta$ values were otherwise fixed for the fits shown. The drived diameters (D [km units]) and albedos ($p_v$) are also shown in the lower right of each of these panels for each model fit. Panel E shows the black-body fit to P/2010 R2, appropriate for a dust-coma dominated signal, through no apparent extended profile was found in the stacked image (see text).

**Figure 4** Coma temperature fit to the 4.3 arcmin aperture thermal photometry in the two longest WISE wavelength bands of the P/2010 A2 observations. A reflected-light model with a neutral reflectance (heavy dotted) based on the near-simultaneous photometry from Jewitt *et al.* (2011a) is shown along with the combined signal (dashed line). The uncertainties to the temperature fit are on the order of +/- 9K, and the fitted temperature (238K) is in excess of the black body temperature for that distance (200K).

**Figure 5** Particle Mass Distribution (PMD) of P/2010 A2. Log number is shown vertically, while log mass is shown on the bottom scale and the corresponding grain radius size, in microns, is shown on the scale above. The P/2010 A2 data derived number of particles in the 4.3 arcmin aperture radius (stars), encompassing the complete signal from the dust tail, are shown. For comparison, 103P/Hartley 2 (pentagons; Bauer *et al.* 2011), Deep Impact particle densities (triangles; Schleicher *et al.* 2006, Bauer *et al.* 2007, and Lisse *et al.* 2007), and Echeclus particle numbers (diamonds and squares; Bauer *et al.* 2008) are also shown. Stardust PMD slope ($\alpha = -0.75$, in log N/log kg units, where N is the estimated total number of dust grains in the aperture; Green *et al.* 2004) is shown as the dashed line, rescaled from dust fluence values to an aperture encompassing a similar $\rho$ size. Echeclus' PMD best-fit ($\alpha = -0.87$) is shown as a dotted line, and the solid line is the best fit to 103P PMD data ($\alpha = -0.97$). The dot-dashed line represents the best fit to the P/2010 A2 data of $\alpha = -0.5$.









**Figures**

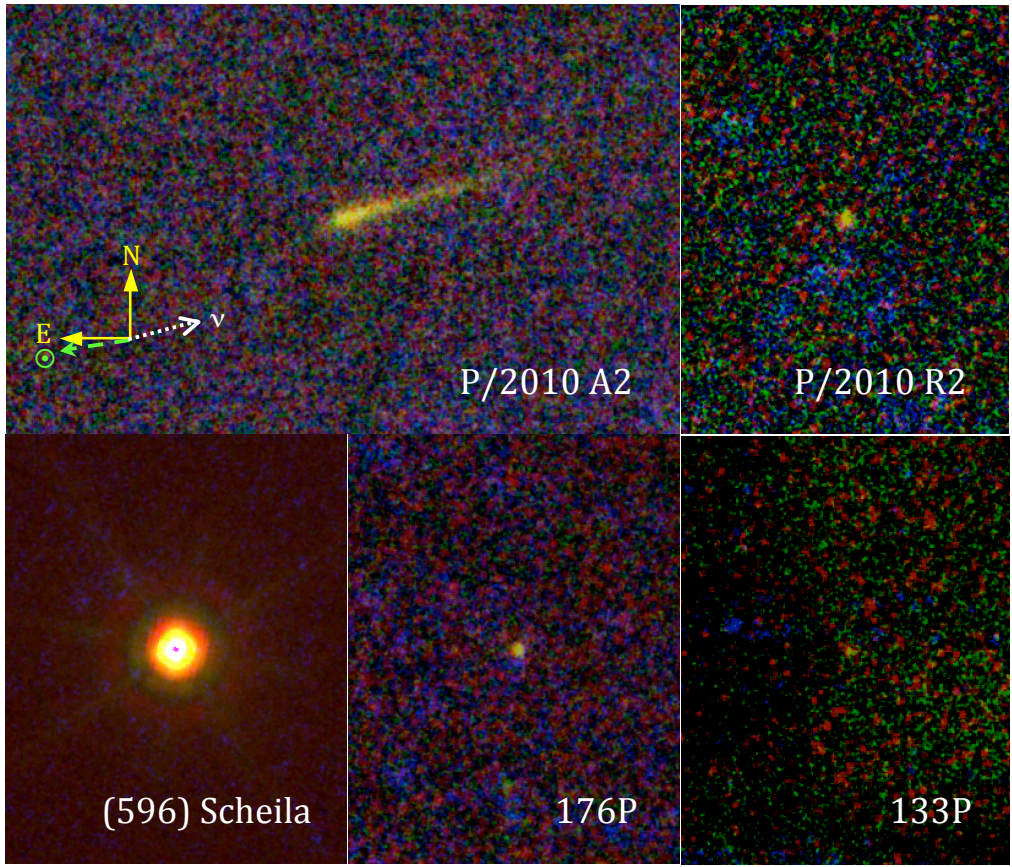

Figure 1





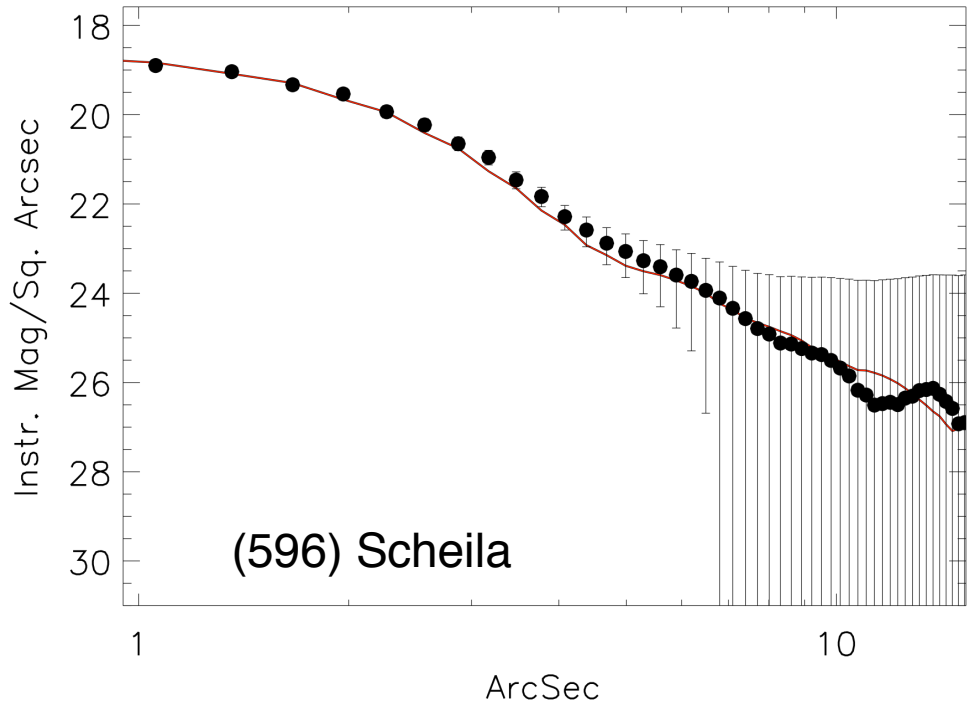

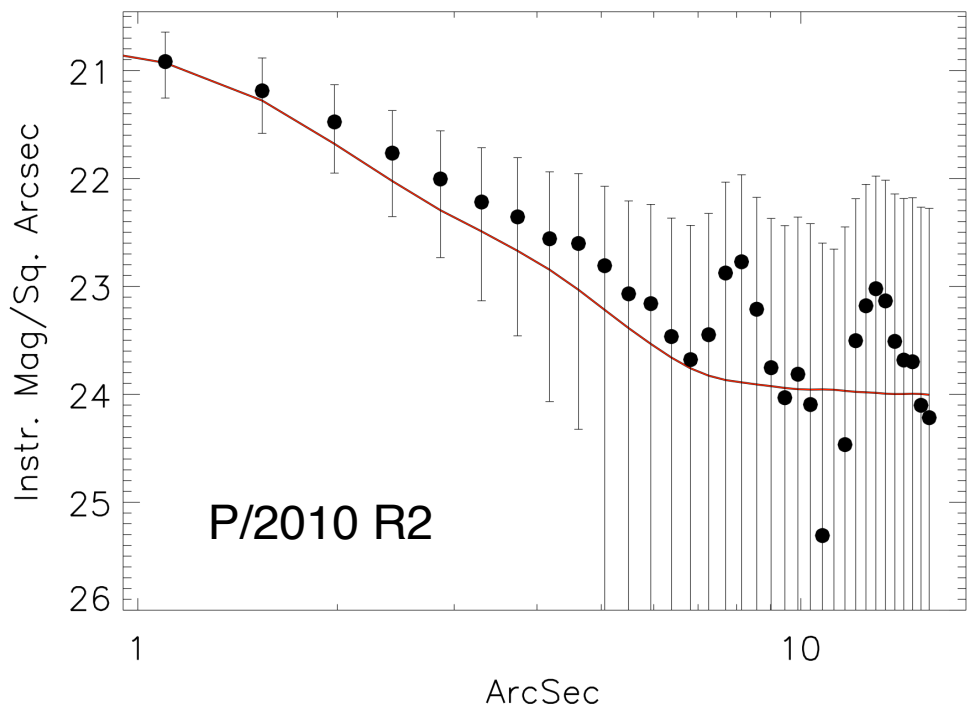

Figure 2





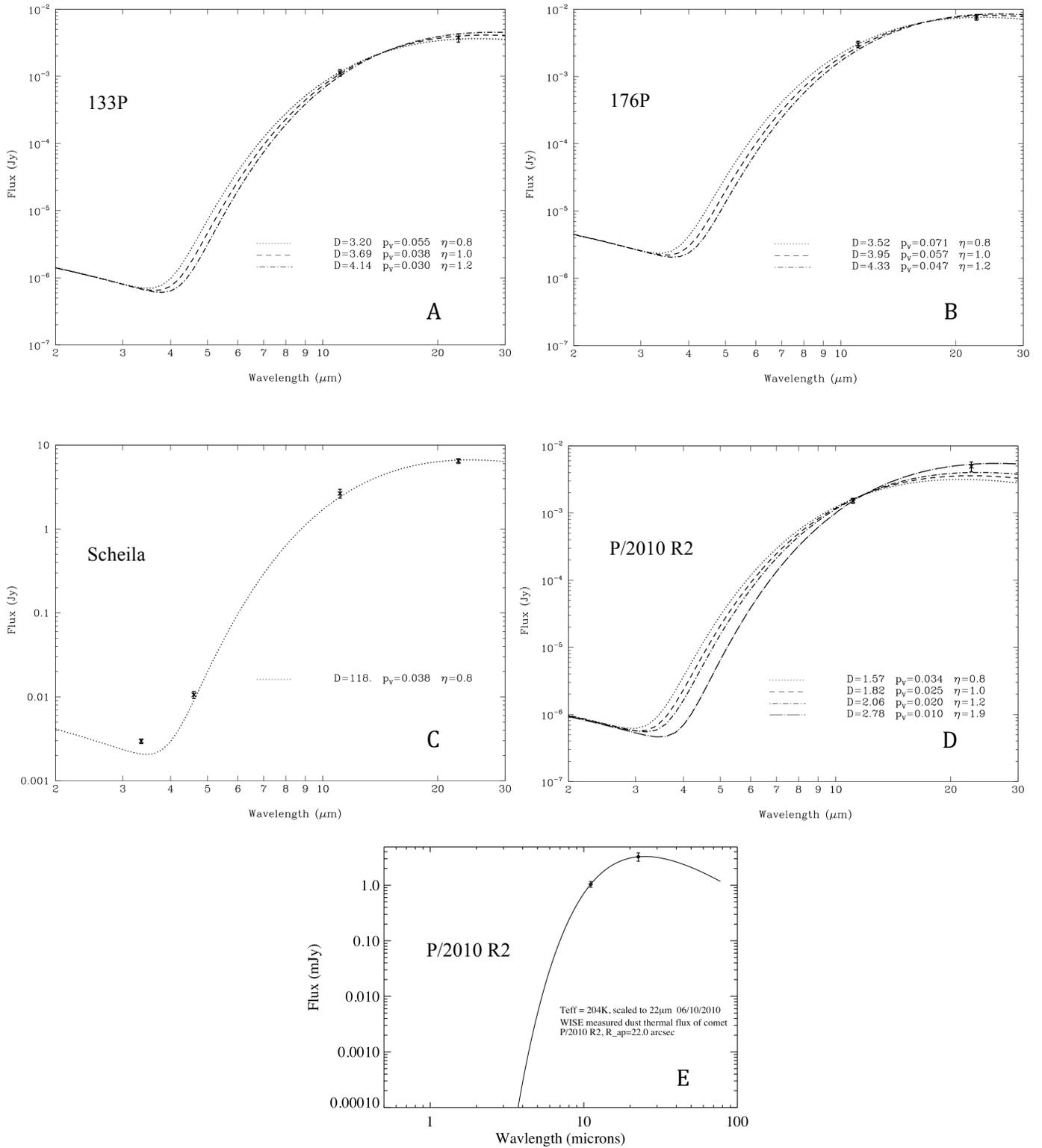

Figure 3





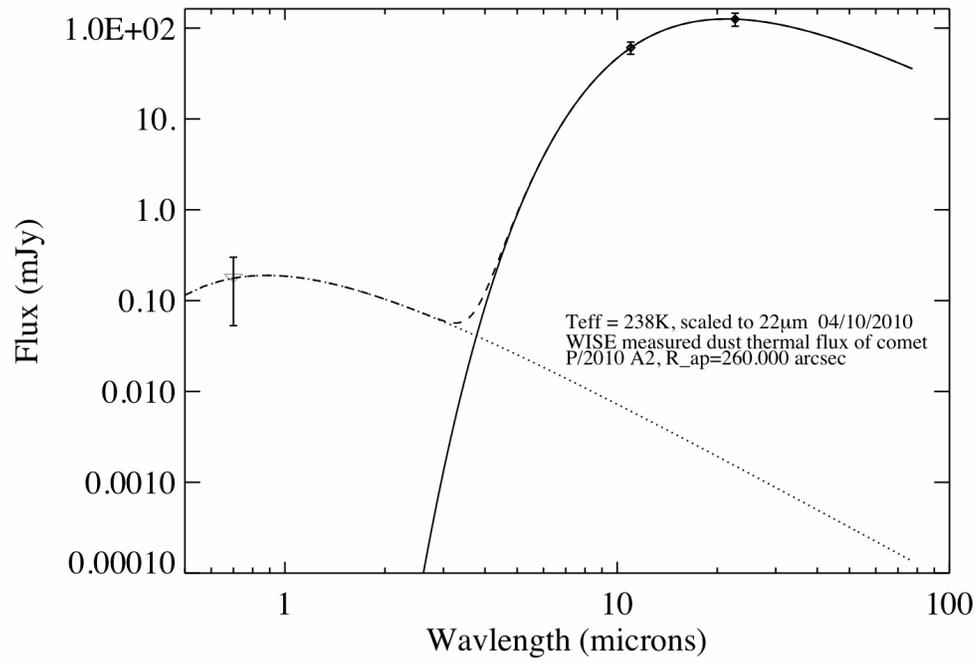

Figure 4





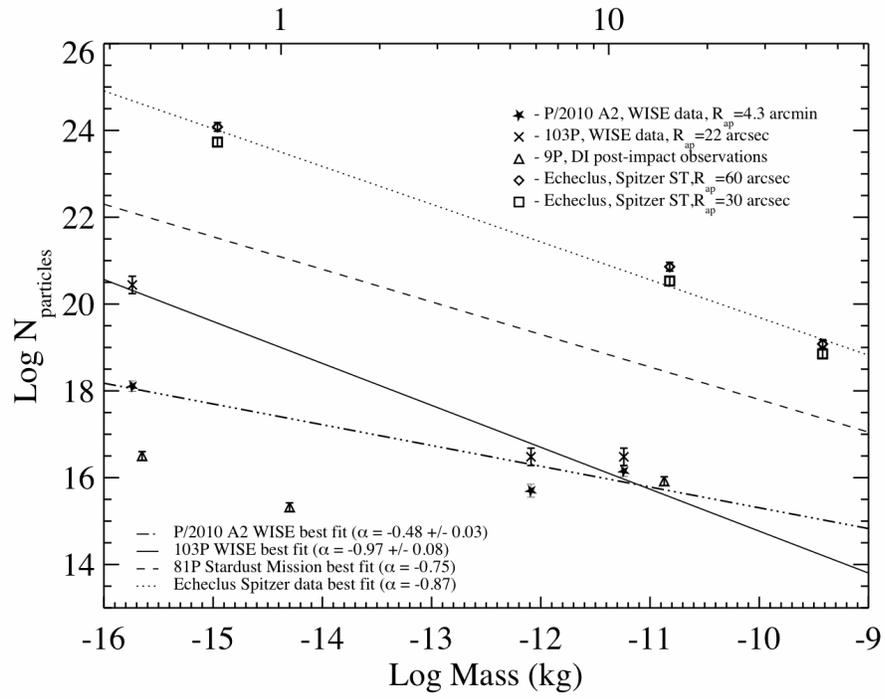

Figure 5